\pdfoutput=1
\documentclass[aps,prl,twocolumn,showpacs,amsmath,amssymb,superscriptaddress,floatfix]{revtex4}
\usepackage{graphicx,epsf}
\usepackage{bm}

\begin{document}

\title{Tunable resonances due to vacancies in graphene nanoribbons}

\author{D. A. Bahamon}
\address{Instituto de F\'\i sica, Universidade Estadual de Campinas - UNICAMP,  C.P. 6165, 13083-970, Campinas, SP, Brazil}
\author{A. L. C. Pereira}
\address{Faculdade de Ci\^encias Aplicadas, Universidade Estadual de Campinas - UNICAMP, Limeira, SP, Brazil}
\author{P. A. Schulz}
\address{Instituto de F\'\i sica, Universidade Estadual de Campinas - UNICAMP,  C.P. 6165, 13083-970, Campinas, SP, Brazil}

\date{\today}

\begin{abstract} 
The coherent electron transport along zigzag and metallic armchair graphene nanoribbons in the presence of one or two vacancies is investigated. Having in mind atomic scale tunability of the conductance fingerprints, the primary focus is on the effect of the distance to the edges and inter vacancies spacing. An involved interplay of vacancies sublattice location and nanoribbon edge termination, together with the spacing parameters lead to a wide conductance resonance line shape modification. Turning on a magnetic field introduces a new length scale  that unveils counter-intuitive aspects of the interplay between purely geometric aspects of the system and the underlying atomic scale nature of graphene.
\end{abstract}

\pacs{73.23.-b, 72.80.Vp, 81.05.ue}

\maketitle


\section{I. Introduction}
Since the experimental realization of graphene \cite{Novoselov1, Novoselov2, Kim1}, the promises of their unique electronic properties \cite{rev_mod_phys} lead to intense efforts to create high quality samples \cite{EAndrei,Kim2} that have permitted to achieve the ballistic regime in electron transport measurements even at room  temperatures. As the quality is increased and characteristic device dimensions are reduced \cite{Wang} graphene is set to become one of potential materials for future electronic devices.
Graphene nanoribbons can be considered as the starting point structures out of which devices may be cast out  for envisaging graphene based electronics. An example, among others are the quantum dots revealing single electron charging signatures \cite{ensslin}. In this context, there is a continuously growing number of proposals for device configurations \cite{Nanomesh,Yamamoto}, including antidot superlattices \cite{jauho,Latge,Peeters,vanevic}. However, the minimal device one can think of is a single atomic vacancy \cite{Deretzis}, acting as the simplest possible antidot in a nanoribbon. The natural subsequent step is a two vacancies (antidots) molecule drilled in a nanoribbon. Such a simple system is expected to reveal already rich transport fingerprints, associated, for instance, to the coupling and decoupling between the antidots and edges by means of tuning a magnetic field or due to quasi bound states in an antidot defined potential landscape, resembling conventional two dimensional devices \cite{Gudmundsson, Satanin}. 
The scenario of structured two dimensional electron gases becomes more involved in the case of graphene due to the new degrees of freedom given by the peculiar features of the carbon atoms monolayer, sensible to engineering on the atomic scale \cite{Mucciolo,Peeters2}. The presence of two sublattices of the honeycomb structure and the definition of two clearly different atomic orientations for the edges of a finite graphene nanoribbon, namely zigzaglike and armchairlike \cite{Nakada}, underline emergent design possibilities of electronic properties for this system, in respect to conventional two dimensional systems. The aim of the present work is to explore the transport fingerprints of the interplay between geometric aspects of tailoring a two dimensional layer and those allied to properties specific to graphene manipulated on the atomic scale. The most appealing findings are the possibilities of tuning the coupling between localized states around the vacancies and continuum states at the edges, a scenario further enriched by a new length scale introduced by turning on a magnetic field. 
We define a possible imaging of the mentioned coupling strength, which characterizes the line shape of the well known Fano resonances, originally discussed for atomic systems almost fifty years ago \cite{Fano}. On the other hand, tuning of Fano resonances in solid state systems became feasible and could be considered as a spectroscopy tool of the underlying electronic structure in different classes of electronic transport devices characterized by at least two channels that may interfere with varying coupling strengths. Initially elusive in mesoscopic systems from an experimental point of view, Fano resonances have been object of comprehensive theoretical and experimental scrutiny \cite{Mendoza,wulf}.
In Section II some basic aspects of the tight binding approach and the lattice Green's-function method are mentioned. In Section III the conductance and local density of state are calculated for zigzag and armchair nanoribbons drilled by single vacancies. In section IV the conductance and local density of state (LDOS) are calculated for zigzag and armchair nanoribbons with two vacancies, discussing the effects of the sublattices of the missing atoms and the separation of the vacancies with and without magnetic field.  Section V outlines the conclusions.

\begin{figure}[t]
\vspace{-0.2cm}
\centerline{\includegraphics[width=8.9cm]{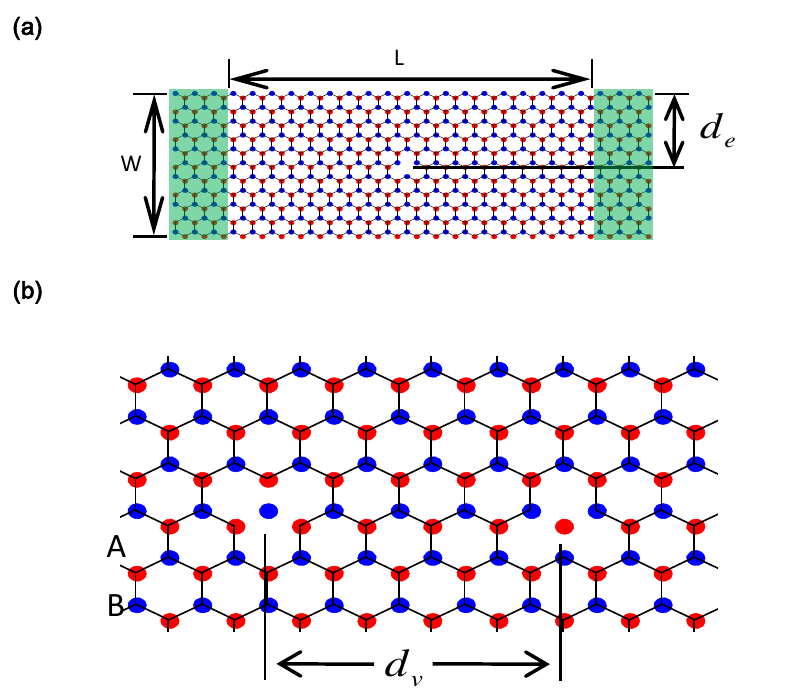}}
\vspace{-0.2cm}
\caption{ (Color Online) {\bf (a)} Schematic representation of a zigzag graphene nanoribbon of length $L$, width $W$ and a vacancy at a distance $d_e$ from the upper edge. {\bf (b)} Geometry of two vacancies separated by a distance $d_v$ in a zigzag nanoribbon: hopping parameters are set to zero to simulate the absent atoms. In this case vacancies are on different sublattices (different colors indicate different sublattices).}
\end{figure}


\section{II. Model}

An infinite zigzag or armchair nanoribbon of width $W$ can be separated in three parts: a central region (device) of length $L$ (in the present work we fix $L=50$ nm) which contains the vacancies embedded  between two perfect graphene ribbons acting as semi-infinite contacts (left and right). We model the electronic structure by a nearest neighbor tight-binding Hamiltonian.

\begin{equation}
H = \sum_{i} \varepsilon_{i} c_{i}^{\dagger} c_{i}
+ t  \sum_{<i,j>} (e^{i\phi_{ij}} c_{i}^{\dagger} c_{j} + e^{-i\phi_{ij}}
c_{j}^{\dagger} c_{i})
\end{equation}

\hspace{-\parindent}where $c_{i}$ is the fermionic operator on site $i$.
The perpendicular applied magnetic field is included by means of Peierls substitution, which means a complex phase in the hopping parameter $t$: $\phi_{ij}= 2\pi(e/h) \int_{j}^{i} \mathbf{A} \! \cdot \! d \mathbf{l} \;$. In the Landau gauge $\mathbf{A} =(-By,0)$ , $\phi_{ij}\!=\!0$ along the $y$ direction and $\phi_{ij}\!=\mp  2\pi (y/\sqrt{3}a) \Phi / \Phi_{0}$
along the $\pm x$ direction for a zigzag nanoribbon and $\phi_{ij}\!=\mp  2\pi (y/a) \Phi / \Phi_{0}$ along the $\pm x$ for armchair nanoribbon, with the magnetic flux per magnetic flux quantum defined as: $\Phi / \Phi_{0}=Ba^{2}\sqrt{3}e/(2h)$, where  $a$=2.46{\AA} the lattice constant for graphene. The magnetic field we consider permeates the entire structure, including the contacts. Vacancies are defined by setting the hopping parameters to zero for the absent atoms and the on-site energies of these atoms equal to a large value outside the energy range of the spectra. Relaxation and reconstructions around vacancies are not considered here, however they should not modify qualitatively the results for the conductance shown in this work \cite{Ana}.  The main ingredient here is that vacancies are localized defects showing states strongly localized around them, as expected from numerical simulations \cite{castroneto2008,lee}. In Fig. 1a the geometry of the central region of width $W$ and length $L$ of a zigzag nanoribbon with a vacancy located at a distance $d_e$ from the upper edge is shown. The different colors for the carbon atoms indicate the two distinct sublattices, A and B. In Fig. 1b, the structure of a two vacancies set, located at different sublattices (B-A), separated by a distance $d_v$ in a zigzag nanoribbon, is represented. In a similar way, for the armchair nanoribbons we have defined a central region of width $W$ and length $L$. For both types of edge termination when considering only one vacancy, it is located at the middle length $(L/2)$, while for two vacancies it is at $(L/2 \pm d_v/2)$. The  diversity of the two vacancies cases is expressed by the sublattice position of the vacant atoms (A-A, A-B, B-A or B-B).

The conductance is evaluated within the Landauer-B$\ddot{u}$ttiker formalism, $G(E)=G_oT(E)$, where $G_o=\frac{2e^2}{h}$ is the conductance quantum  and $T(E)$ is the transmission function between the contacts and can be calculated by \cite{Datta}: 

\begin{equation}
T_{pq}=Tr[\Gamma_pG^r_{pq}\Gamma_qG^\dagger_{pq}]
\end{equation}

\hspace{-\parindent}where $G^r_{pq}$ is the Green's function between the contact $p$ and $q$ and is evaluated thorough the recursive lattice Green's function technique \cite{PALee, Ferry}. This numerical protocol has been extensively applied to systems described by square lattice models, here adapted to the honeycomb geometry \cite{Mucciolo}. The contact broadening function \cite{cresti1} $\Gamma_{p(q)}=i(\Sigma_{p(q)}-\Sigma^\dagger_{p(q)})$ where $\Sigma_{p(q)}$ is the self-energy of the contact, arises from the interaction of the central region with the semi infinite contact. In order to calculate this term, it is necessary to know the Green's function of the contact, also obtained numerically \cite{LSancho}. The local density of states (LDOS) at a given site $j$ can be straightforwardly extracted from   $\rho_j=-\frac{1}{\pi}Im[G_{jj}]$ where $G_{jj}$ is the total Green function at site $j$ \cite{Mendoza}.

\section{III. SINGLE VACANCIES: discrete energy states embedded in a continuum}

A throughout description of the problem is, as stated in the abstract, an involved interplay of vacancies sublattice location, inter-vacancy separation, distances to the edges, as well as the edge geometry. In view of this complexity we start by establishing an appropriate framework of analysis, which will consist in a sublattice resolved local density of states description that will be linked to conductance resonance line shapes. In order to build up such framework, we initially focus on the intensively studied single vacancy problem.

When a vacancy is created in graphene, a zero energy state appears, with the corresponding wave function strongly localized around the missing atom and only on one of the sublattices \cite{VPereira}. In a zigzag nanoribbon the overlap between  the edge states leads to one excess channel with a defined chirality \cite{Brey1}, therefore for this channel the vacancy state located on an A(B) sublattice couples with the A(B) edge state, leading to conductance dips at the corresponding energies of the bonding and antibonding configurations \cite{Wakabayashi2}. 

Fig. 2a  shows the conductance (in units of $G_o=2e^2/h$) as a function of Fermi energy (parametrized by the hopping parameter $t$) for a zigzag nanoribbon with $W=15.9$ nm, containing one vacancy at three different positions. The continuous line is for a vacancy located either on A or B sublattice at $d_e = W/2 = 8$ nm, i.e. in the middle the nanoribbon. Hence, the couplings with an A(B) sublattice terminated lower(upper) edge of the nanoribbon  are the same and a wide dip at $E=0$ is observed.
The dashed line in Fig. 2a corresponds to an A sublattice vacancy at $d_e=4$ nm (i.e., significantly closer to the upper edge). The absence of noticeable signatures on the conductance is understood as follows. A vacancy on an A sublattice creates, as already mentioned, a state localized solely on B sublattice sites. Furthermore, since the upper edge is terminated by B sublattice sites, the density of states is predominantly also on the B sublattice in the region between the edge and the vacancy. Therefore, no sublattice mixing effects are expected that could trigger changes in the conductance. On the other hand, changing the vacancy sublattice (now a B vacant site), but keeping the same distance, $d_e$, to the edge, a rather dramatic change in the conductance is observed, as can be observed in the dotted line of Fig. 2a, where a dip at a nonzero energy is observed, in agreement with Ref. \cite{Wakabayashi2}. 

Figure 2c maps the LDOS at the energy of the conductance dip of the dotted line case in Fig. 2a, while Fig. 2d pictures a complementary view (contour plot) of the same LDOS. The LDOS is color differentiated for both sublattices, A(red) and B(blue). Clearly there is a high LDOS peak in the A sublattice near the B vacant site that decays as the distance to the vacancy increases and with a definite symmetry \cite{Rutter}, as expected. Noticeable here is a significant LDOS modulation on the B sublattice between the upper edge (clearly identifiable on Figs. 2c and 2d as a bump in the LDOS at the upper edge) and the vacancy, indicating a rather strong edge-vacancy coupling, as will be discussed below.

\begin{figure}[t]
\vspace{-0.2cm}
\centerline{\includegraphics[width=8.5cm]{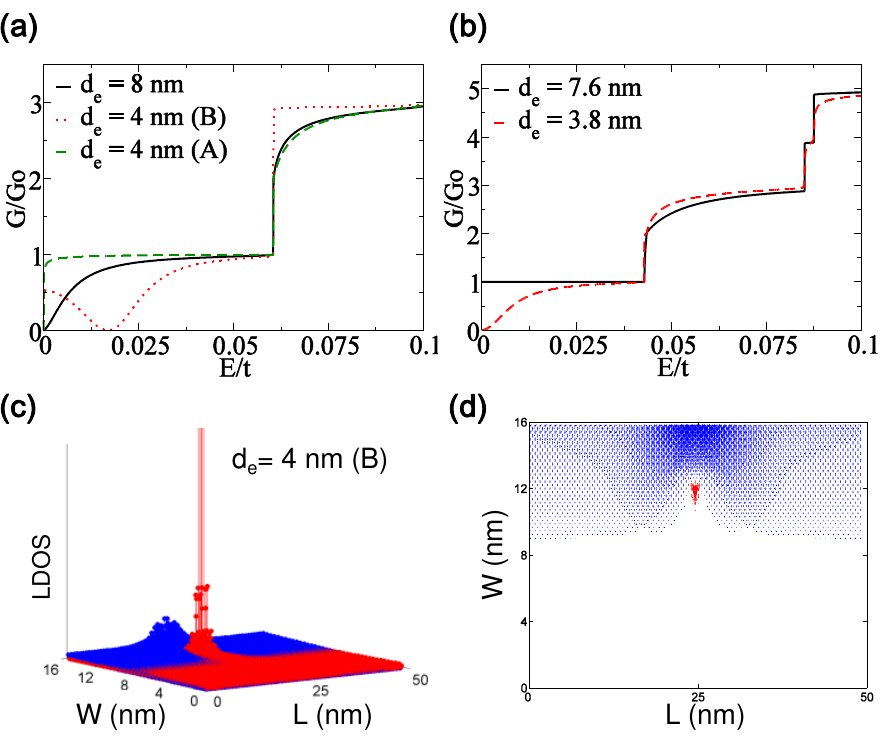}}
\vspace{-0.2cm}
\caption{(Color Online) {\bf (a)} Conductance for a zigzag nanoribbon of $W=15.9$ nm with  a vacancy at different  distances from the upper edge ($d_e$) and at different sublattices. {\bf (b)} Conductance for a metallic armchair nanoribbon of $W=15.2$ nm with a vacancy at different  distances from the upper edge ($d_e$). {\bf (c)} LDOS, color differentiated for both sublattices, A and B, corresponding to the zero of conductance of the dotted line of Fig. 2a (B vacancy site at $d_e=$ 4 nm). {\bf (d)} Contour plot of the same LDOS depicted in (c).  A highly localized state is observed on the A sublattice sites around the vacancy and also a modulation on the LDOS of the B sublattice sites between the vacancy and the closer (upper) edge.} 
\end{figure}


The effect of a vacancy in a metallic armchair ($W=15.2$ nm) is shown in Fig. 2b. For armchair nanoribbons, the main changes in the line shape of the conductance are due to the variation of distance to the edges, $d_e$, and not to the vacancy sublattice. The independence of vacancy sublattice for armchair edges is expected due to equal participation of both sublattices on the edge termination. Here, the continuous line is for either an A or B vacancy in the middle of the nanoribbon ($d_e=7.6$ nm) and no effect in the first conductance channel is seen. On the other hand, for either an A or B vacancy closer to one of the edges, at $d_e=3.8$ nm, a dip at the Dirac point develops (dashed line in Fig. 2b). Indeed, moving the vacancy from the center to the edge of the nanoribbon lead to systems that alternate periodically between perfect conducting channels and zero energy dips \cite{Xiong}. 

For both type of edge termination, the effect of the vacancy sublattice or distance to the edge ($d_e$) slightly affect higher energy channels as can be expected from the scattering in a single repulsive potential.

These results, as well as the following ones for double vacancies, should be seen in a broad context of discrete energy states embedded in a continuum, hence two alternative quantum pathways interfere leading to resonances with  different Fano line shapes \cite{Fano}. These line shapes are characterized by a Fano asymmetry factor $q$, a measure of the coupling strength between the continuum and the resonance state (which is a modification of the discrete state through the interaction with the continuum).  For $q=0$, one observes a perfectly symmetric anti-resonance, i.e, a strong dip in the amplitude of $G_0$ is found at the energy of the resonance. Finite $q$ leads initially to asymmetric resonance profiles, the well known dip-peak structure in the conductance and for $q \rightarrow \infty$ a symmetric Breit-Wigner-type resonance profile is recovered.  

The present results suggest an image for the $q$ strength of Fano type resonances. Recalling that the anti resonance in the dashed line curve in Fig 2a is not completely symmetric, a finite $q$ should indicate a finite coupling between the localized state (associated to the vacancy) and the delocalized continuum (edge states). Indeed, in Fig. 2c the bump at the edge LDOS is due to the presence of an off center vacancy at the same sublattice of the edge.

\section{IV. DOUBLE VACANCIES}

When two vacancies are introduced in a graphene sheet, the effect of the sublattice becomes rather crucial. In the case of both vacancies belonging to the same sublattice, two zero energy modes will be created \cite{VPereira}. On the other hand, if they are located on different sublattices states away of the Dirac point appear \cite{VPereira}. In nanoribbons an additional ingredient seems to be relevant, in order to characterize the transport properties of such devices: besides the distances between vacancies ($d_v$) \cite{Ana,Ma} and sublattice assignment of the vacancies \cite{Palacios}, the proximity to an edge, $d_e$, also plays an important role. It should be kept in mind that adding a new variable to the problem may lead to emergent characteristics, as will be seen here in the manifold of resonance line shapes.

\subsection{A. Conductance line shapes: Effects of coupling between vacancies}

We initially change $d_v$,  creating therefore different types of double vacancies: A-A, B-B, B-A  or A-B (letters indicate the sublattice where each vacancy is located), in absence of magnetic field.


\begin{figure}[b]
\includegraphics[width=8.7cm]{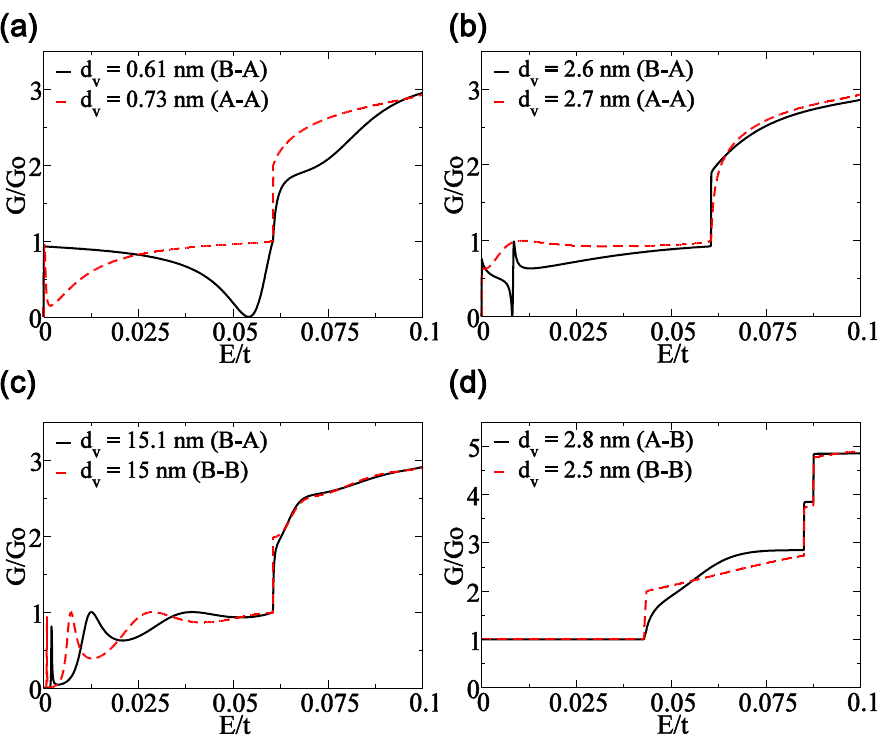}
\vspace{-0.3cm}
\caption{ (Color online) {\bf (a)}, {\bf (b)} and {\bf (c)} Conductance for a zigzag nanoribbon ($W=15.9$ nm) with two vacancies, both at the middle width of the ribbon ($d_e= W/2=$ 8 nm), varying the distance ($d_v$) between vacancies and vacancy sublattices. {\bf (d)} Conductance for a metallic armchair nanoribbon ($W=15.2$ nm) with two vacancies both at  $d_e=$ 7.6 nm, varying the sublattice assignment.}
\end{figure}


For a zigzag nanoribbon of $W=15.9$ nm we start by considering $d_e=W/2$, i.e., the vacancies at the middle of the ribbon. The smallest distance between vacancies considered is $d_v=0.61$ nm, determining a B-A like double vacancy. The resulting conductance is given by the continuous line in Fig. 3a, where an anti-resonance near the first conductance step appears as a result of a high coupling of both vacancies that belong to different sublattices, generating a bonding and antibonding energy states scenario. This strong coupling effect also changes the line shape of the higher conductance steps. If we take the vacancy located at the B sublattice and move it to its left nearest neighbor an A-A double vacancy is obtained with a $d_v=0.73$ nm. The conductance of this configuration is the dashed line in Fig. 3a. Such subtle change in the position of an absent atom drastically modifies the conductance line shape, where now a dip in conductance appears for an energy near to the Dirac point although the conductance never goes exactly to zero. Also noticeable is the conductance enhancement at higher steps. Effects of increasing the distance between both vacancies can be inspected in Fig. 3b: the continuous line of Fig. 3b corresponds to a B-A double vacancy with $d_v=2.6$ nm (an intermediate coupling between the vacancies). The formation of a typical asymmetric Fano line shape in the first channel is evident, manifesting the quantum interference feature of the transmission through the continuum and the resonant states ($|q| \approx 1$). Comparing this line shape with the dashed line in the same figure, namely for a A-A double vacancy with $d_v=2.7$ nm, we observe that now the asymmetric Fano profile is washed out, evolving a broad featureless conductance profile. 
Increasing the distance between vacancies will decrease their coupling, as expected for B-A double vacancy with $d_v=15.1$ nm, a case with the corresponding conductance plotted as a continuous line in Fig. 3c. Now the transmission resonances due to the presence of double vacancies are bona fide symmetric Breit-Wigner-type peaks, ($|q|=\infty$), typical of resonant tunneling structures, where the continuum background is strongly suppressed at the discrete state localization. The dashed line in Fig. 3c shows the conductance of a B-B double vacancy with $d_v=15$ nm, where the same Breit-Wigner resonance appears, but shifted to lower energies. Both line shapes in Fig. 3c are similar, since the inter-vacancy distance in both cases are quite large and the coupling is mainly between each vacancy and the edge. The coupling of both vacancies to the edges transforms the otherwise separated defects into a unique scattering center with resonances resembling the ones of a double barrier 
 resonant tunneling structure \cite{Datta}. 

 For a corresponding armchair nanoribbon ($W=15.2$ nm $d_e=W/2$), depicted in Fig. 3d, it is noticed that the conductance for either an A-B or B-B double vacancies, with $d_v=2.8$ nm (continuous line)  $d_v=2.5$ nm (dashed line), respectively, the first conductance channel is not modified respective to the bare nanoribbons. It should be mentioned that A-B and B-B double vacancies show no differences in respect to B-A and A-A arrangements, either in zigzag or armchair nanoribbons, whenever the vacancies are in the middle of the ribbon.  

\subsection{B. LDOS mapping: Coupling of vacancies and edges}

We showed therefore that the presence of double vacancies on different sublattices in zigzag nanoribbons gives origin to resonances in the conductance, which have line shapes that depend on the distance between vacancies. In order to correlate these line shapes with the tuning of the coupling between vacancies and between vacancies and edges, sublattice color differentiated LDOS are depicted in Fig. 4, recalling that A(B)-like sites are identified by red(blue). 
In Fig. 4a, at the energy of the anti-resonance generated by the B-A double vacancy with $d_v=0.61$ nm (corresponding to the continuous line in Fig. 3a), two close peaks, each at distinct sublattices are clearly identified, but no noticeable change on the background LDOS at the edges. Such signature corresponds to a weak continuum to localized state coupling, $q=0$. Fig. 4b shows the mapping of LDOS for the energy of zero conductance in the asymmetric line of the B-A double vacancy with $d_v=2.6$ nm (continuous line in Fig. 3b). Here also two peaks at different sublattices appear, with the LDOS peak at B sublattice (blue) evolves in a tail extended to the upper edge, while the LDOS peak at A sublattice (red) evolves into a tail extended to the lower edge, showing a clear example of how a discrete state is modified by the continuum. On the other hand, the bumps at the edge LDOS correspond to a finite value of $q$, fingerprint of the vacancy-edge coupling leading to an asymmetric Fano profile. The LDOS of the first Breit-Wigner peak for the B-A double vacancy, observed in Fig. 3c, is also mapped in Fig. 4c: two well separated peaks at different sublattices are highly coupled to the upper and lower edge, respectively on the B and A sublattices. For the B-B double vacancy case, Fig. 4d, two peaks on the A sublattice appear highly coupled to the lower edge. Furthermore an additional high density is located at the upper edge, with a maximum value in the middle of the device, evidencing also a strong coupling to the upper edge. From the last two figures, Fig. 4c and 4d, it is evident that the two vacancies, no matter their sublattices, induce a strong modification on the continuum, i.e., $q \rightarrow \infty$. In other words, the height of the bump in the LDOS at the edges is proportional to the absolute value of $q$.
Background LDOS at the edge compared to the enhancement maximum in the vacancy region.

\begin{figure}[t]
\includegraphics[width=8.4cm]{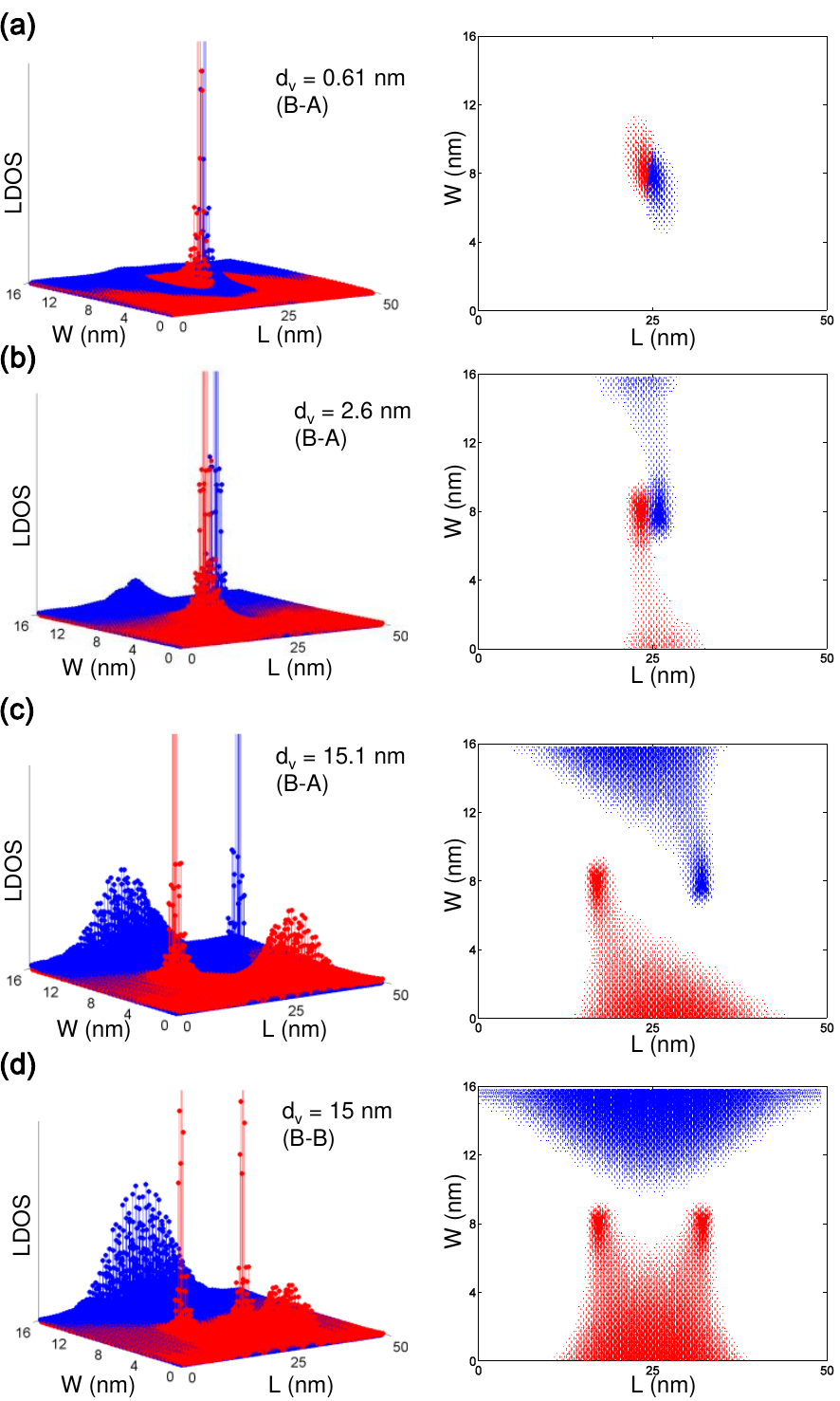}
\caption{ (Color Online) LDOS (left panels: 3D plot; right panels: contour plot) of the  {\bf (a)} zero of conductance  for the continuous line of Fig. 3a. {\bf (b)} zero of conductance  for the continuous line of Fig. 3b. {\bf (c)} first Breit-Wigner peak of conductance  for the continuous line of Fig. 3c. {\bf (d)}  first Breit-Wigner peak of conductance  for the dashed line of Fig. 3c. Around vacancies, the LDOS, which is color differentiated for both sublattices, A and B,  concentrates on the oposite sublattice from the vacant atoms.  The relations of line shapes from Fig.3 with couplings between vacancies and edges are evident.}
\end{figure}

All the results shown above for double vacancies have the distance to the edges fixed $d_e=W/2=8$nm.  Considering the B-A double vacancy with $d_e=2.6$ nm, which show a resonance with an asymmetric line shape, a translation to $d_e=W/4=4$nm should change the resonance fingerprints on the conductance. This case is investigated in Fig. 5. The overlap between the B vacancy and the upper edge is increased, while the overlap between the A vacancy and the lower edge is reduced but neither would be equal to the line shape of a single vacancy at the same $d_e$ because of the inter-vacancy coupling. In Fig. 5a the conductance for the new (dashed line) and for the previous $d_e$ (continuous line) are depicted for comparison. Proximity to the upper edge broadens the resonance and it should be noticed that the corresponding single vacancy case, Fig. 2a (dotted line), is indeed still different, evidencing the coupling to the second vacancy in the present case. Further evidence is presented in Fig. 5b: the LDOS at the energy of conductance dip of the dashed line case is plotted and a high LDOS peak in the A sublattice (red) is observed (with the already mentioned coupling to the upper edge), while the LDOS around the position of the A vacancy is strongly suppressed and the small peak on the B sublattice (blue) LDOS is barely seen (the contour plot shows that this peak on the B sublattice sites merges with the modulation at the LDOS on the upper edge.
The comparison among both double vacancies in Fig.5 and the single vacancy case in Fig. 2c calls the attention to the competition between inter vacancy and vacancy-edge couplings. The double vacancy character is both tuned by the edge distance and the sublattice choice.

\begin{figure}[t]
\begin{center}
\includegraphics[width=8.5cm]{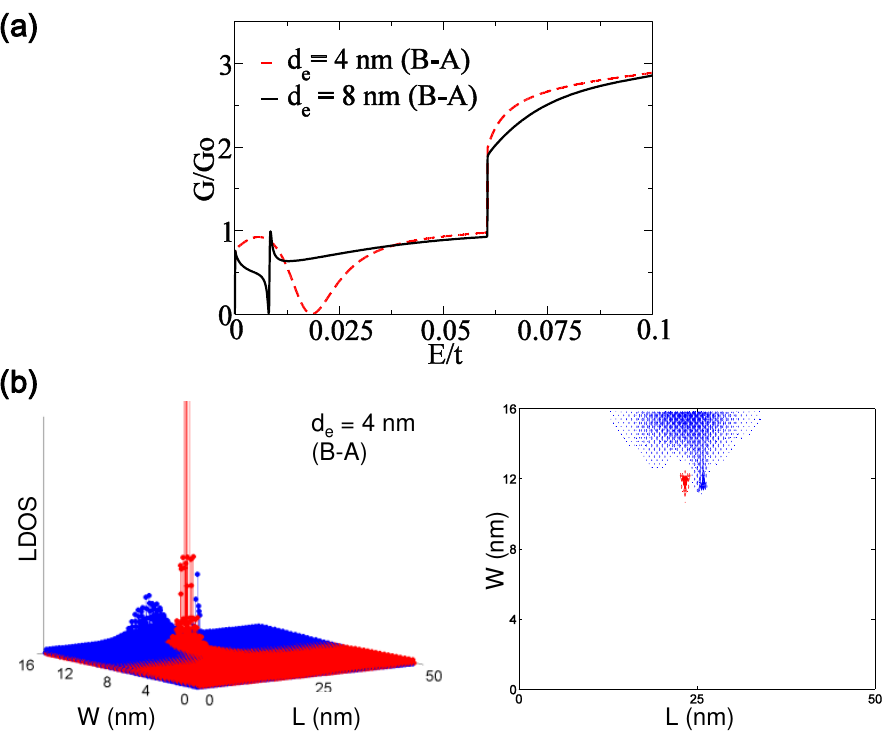}
\vspace{-0.3cm}
\end{center}
\caption{ (color online) {\bf (a)} Conductance for a zigzag nanoribbon ($W=15.9$ nm) with  a B-A like double vacancy at  $d_v= 2.6$ nm and $d_e=W/2=8$ nm (continuous line) and a B-A double vacancy closer to the upper edge, with $d_v= 2.6$ nm and $d_e=W/4=4$ nm (dashed line). {\bf (b )} LDOS corresponding to the zero of conductance of the dashed line of Fig. 5a. (Left panel: 3D plot. Right panel: contour plot.)}
\label{Fig4} \end{figure}

\subsection{C. Manipulating the lengths scale: Double vacancies in a magnetic field}


\begin{figure}[b]
\begin{center}
\includegraphics[width=8.5cm]{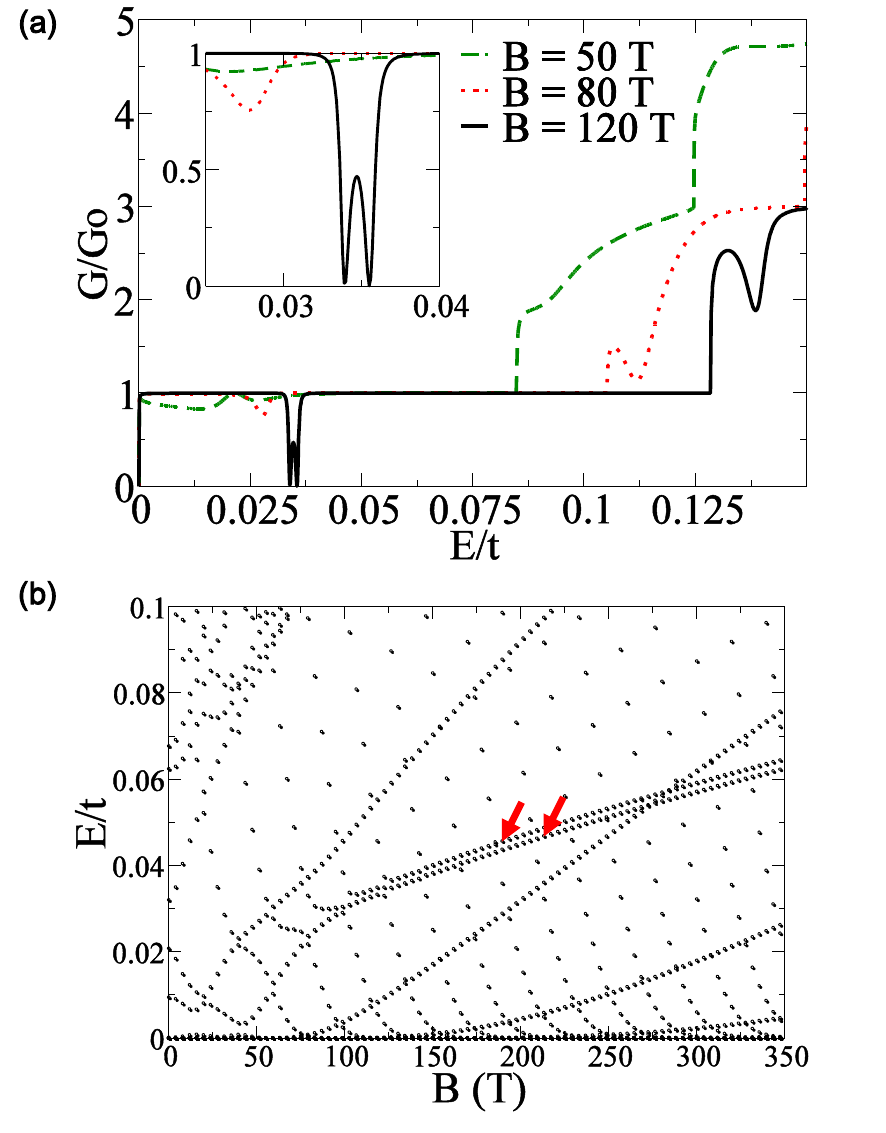}
\vspace{-0.3cm}
\end{center}
\caption{ (Color Online) {\bf (a)} Conductance for a zigzag nanoribbon ($W=15.9$ nm) with a B-A double vacancy ($d_e=W/2=$ 8 nm and $d_v=$ 2.6 nm) for three different magnetic fields. {\bf (b)} Energy spectra as function of the magnetic field for the same nanoribbon, showing the evolution of the two vacancy levels (indicated by the arrows) between the zero and first Landau levels.}
\end{figure}

As we have seen, the appearance of conductance dips, asymmetric resonance line shapes or Breit-Wigner peaks is due to the coupling of two scattering channels, a coupling that may be designed by proper positioning of vacancies. The addition of a magnetic field will introduce new ingredients to the problem: (i) a new length scale, namely the magnetic length ($l_B= \sqrt{\hbar /eB}=25.7 /\sqrt{B(Tesla)}$ nm) and, (ii) an energy shift: the energies of the vacancies states which appear between Landau levels will increase with magnetic field, as investigated both for infinite graphene sheets and finite flakes \cite{Ana,yo}. The effects of applying a magnetic field are scrutinized for the double vacancy systems shown in Fig. 3b. For the B-A double vacancy with $d_v=2.6$ nm, the asymmetric sharp Fano resonance evolves in the presence of magnetic fields, into a wide conductance dip that further evolves as an anti-resonance and finally splits in two anti-resonances, as depicted in Fig.
  6a. Considering that this zigzag nanoribbon is $W=15.9$ nm wide and $d_e=W/2$, turning on the magnetic field smears out the resonances since for very low magnetic fields, $l_B >> W$, both edges and vacancies are mixed up (the new length scale introduced as ingredient (i)). For magnetic fields corresponding to $l_B < W/4$ one expects the recovery of a scenario of vacancies slightly coupled to the edge states and indeed a rather broad asymmetric anti resonance is recovered, but the pathways interference conditions for a sharp Fano like line shape is not fulfilled. We followed the evolution of the anti-resonances that always move to higher energies (the energy shift introduced as ingredient (ii)), and became narrower and eventually disappear at the considered energy scales at extremely high fields, $B\approx 300$ T. This narrowing is consistent with the progressive decoupling between vacancies and edge states with shrinking magnetic length. We also calculated the eigenvalues o
 f such nanoribbon, Fig. 6b, finding that the vacancy localized states have a perfectly match with the energies of the anti-resonances for all the range of magnetic fields where the anti-resonances were observable. In Fig. 6b, the mentioned vacancy states consist of the doublet actually formed for $B >100$ T (indicated by the arrows) and for lower magnetic fields the spectra reveals a so called weak magnetic limit where edge states, Landau levels and localized defect states are not clearly identified.

In Fig. 7a the conductance spectrum for a A-A double vacancy ($d_v=2.7$ nm) for increasing magnetic field is depicted. It is clearly seen that a single dip in the conductance is developed and evolves to a sharp anti-resonance, that apparently never splits into a doublet, as in the B-A double vacancy case.

\begin{figure}[b]
\vspace{-0.2cm}
\centerline{\includegraphics[width=9cm]{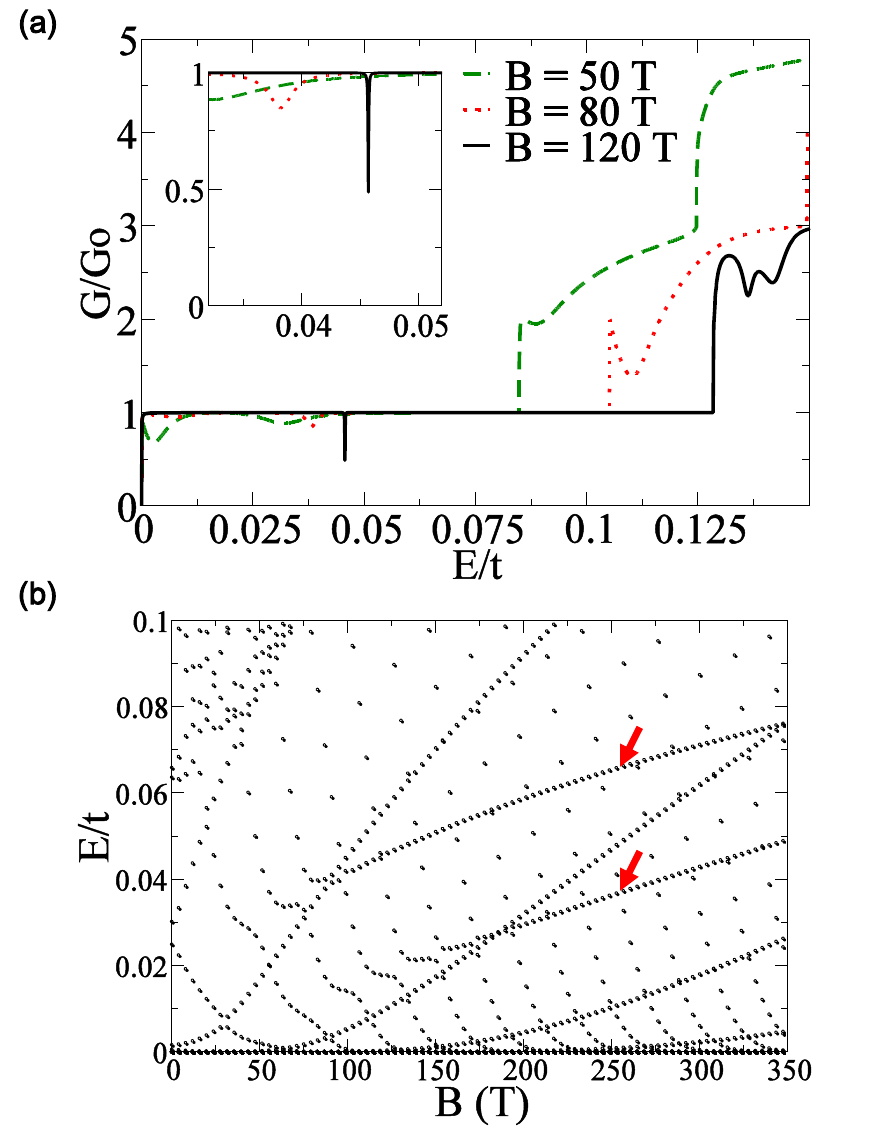}}
\vspace{-0.2cm}
\caption{ (Color Online) {\bf (a)} Conductance for a zigzag nanoribbon ($W=15.9$ nm) with an A-A double vacancy ($d_e=W/2=$ 8 nm and $d_v=$ 2.7 nm) for three different magnetic fields. {\bf (b)} Energy spectra as function of the magnetic field for the same nanoribbon, showing the evolution of the  two vacancy levels (indicated by the arrows) between the zero and first Landau levels. States due to vacancies on the same sublattice show a more pronounced splitting when compared to states due to vacancies on different sublattices (Fig. 6b).}
\end{figure}


Although intriguing at a first sight, the corresponding eigenvalue spectrum in Fig. 7b reveals two important aspects. First, the splitting between the states is more pronounced here (observe the two states indicated by the arrows), since both vacancies are on the same sublattice and therefore a stronger coupling is expected than in the previous case where each vacancy is on a different sublattice. Such subtleties in graphene vacancies electronic structures have been previously investigated in infinite carbon atoms sheets \cite{Ana,Palacios}. Furthermore, the clear second vacancy related eigenvalue is identifiable only for $B> 150$ T, a condition for which the decoupling to the edge is already so strong that the corresponding anti resonance in the conductance has been narrowed down. 
A naive expectation suggests that the splitting between the vacancy states should shrink with increasing magnetic fields, namely for magnetic length smaller than half the inter vacancy distance. Considering the situation depicted in Fig. 6a, $d_v / 2 = 1.3$ nm, representing a magnetic length corresponding to $B = 400$ T.  By inspecting Fig. 6b, no shrinking of the splitting is seen up to $B = 350$ T and indeed this shrinking is significant only for $B > 2500$ T (not shown here), which corresponds to much shorter magnetic lengths, $l_B < 0.5$ nm. This rather unexpected result has also to take into account the quite unique features of electronic states strongly confined to a sublattice in a system where structure changes on the atomic scale may lead to qualitative changes in transport resonance line shapes.

\begin{figure}[t]
\vspace{-0.2cm}
\centerline{\includegraphics[width=9cm]{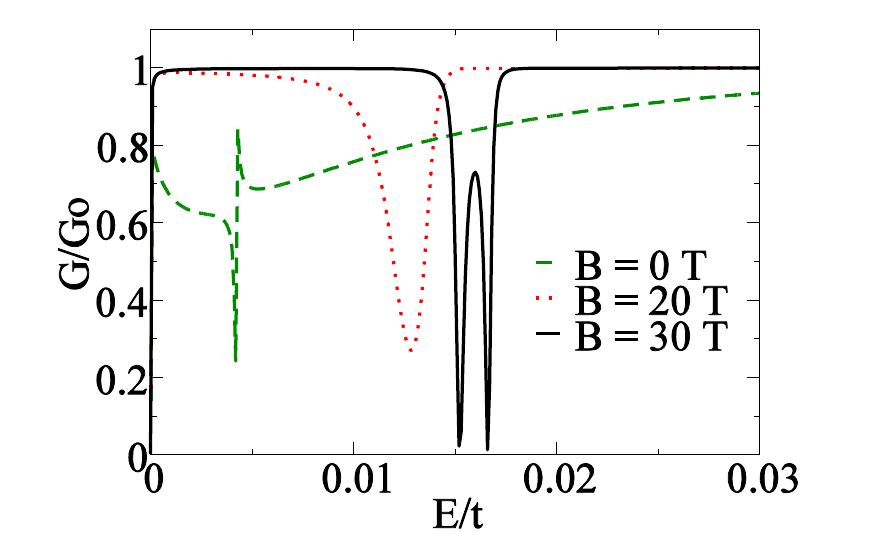}}
\vspace{-0.2cm}
\caption{(color online) Conductance for a wider zigzag nanoribbon ($W = 32.1$ nm) with a B-A double vacancy with $d_v=2.6$nm and $d_e=W/2=16.05$ nm,  for three different magnetic fields. The main features of the evolution of the line shapes are kept when compared to Fig. 6a, however scaled down to lower magnetic fields, making evident the relation between line shapes and edge-vacancies couplings.}
\end{figure}


Looking for more realistic magnetic fields, in Fig.8 we calculated the conductance of a wider zigzag nanroribbon, in order to have a decoupling between the edges and the vacancies at lower magnetic fields. The new zigzag nanoribbon of width $W=32.1$ nm and a B-A double vacancy at the center and with $d_v=2.6$ nm, shows a resonance with an asymmetric line shape in the absence of a magnetic field (green dashed line). It should be noticed that this resonance occurs at lower energies than for a similar device in the narrower ribbons, Fig. 3b, indicating clearly a coupling to the edges effect. Turning on the magnetic field, at  $B\approx 10$ T the conductance fingerprint is washed out (not shown here), but, as can be seen in the dotted line of Fig. 8, at $B=20$ T a dip is developed that evolves into two anti-resonances (continuous line) already at $B=30$ T (analogous to the two anti-resonances observed only after $B\approx 120$ T in the  narrower ribbon at Fig.6a).  Such situation
  should be achievable in an experimental set up, nevertheless the robustness of such effects regarding disorder  \cite{Mucciolo} should be considered in a future work.

\section{V. Final remarks}

In summary, we present conductance spectra for nanoribbons drilled by one or two single vacancies, minimal realization of antidot devices on graphene. Nanoribbons with a single vacancy reveal wide slightly asymmetric anti resonances for a wide range of parameters considered. On the other hand, in nanoribbons with two vacancies (antidot molecule), a manifold of resonance line shapes may occur: anti resonances, Fano-like line shapes and  Breit Wigner conductance peaks, depending subtly from edge termination, sublattice choice for the missing atoms and distances between edges and vacancies ($d_e$) and between vacancies ($d_v$). The coupling between the localized states around the missing atoms is mediated by the coupling to the edges and the resonance line shape can be characterized by the local density of states modulation at the edges. Such modulation could be seen by scanning probe microscopy \cite{westervelt} imaging. 

Furthermore, a perpendicular magnetic field introduces a new length scale to the problem which permits a complete decoupling between edges and antidot related states.  The results shown here further supports the quite unique features of electronic states strongly confined to a sublattice in a system where structure changes on the atomic scale may lead to qualitative changes in transport resonance line shapes.

Fano resonances in graphene systems appear also in more complex potential or geometrical landscapes involving large carbon atom clusters, like in quantum point contacts \cite{wakabayashi3}. Recently, transport properties of graphene sheets with finite concentration of vacancies have been simulated, where the quasilocalization of vacancy related states and band formation are identified \cite{katsnelson}. Nevertheless, the systematic investigation of atomic scale engineering possibilities has to address the scenario of few controlled defects, as the single and double vacancies systems discussed in the present paper. Two general ingredients are of paramount importance in this analysis: the localized character of the vacancy related states\cite{castroneto2008} and,  additionally, as we showed here, the role played by  the sublattices. The latter  should be measurable in future works on the subject, as suggested by recent experiments scrutinizing the sublattice distribution of the electronic density \cite{miller}. 
Reconstructions might influence such pictures, nevertheless first principle calculations indicate that, although relaxations indeed occur around vacancies, bona fide bond reconstructions seem to be absent \cite{iijima,Ma2}.

D.A.B. acknowledges support from CAPES, A.L.C.P. acknowledges support from FAPESP. P.A.S. received partial support from CNPq. Numerical calculations were developed at CENAPAD-SP and IFGW (UNICAMP) computing clusters.


\end{document}